\documentclass[aps,preprint,amssymb,12pt,floatfix]{revtex4}
\pdfoutput=1

\usepackage{subfigure}
\usepackage{graphicx}
\usepackage{lscape,graphicx}
\usepackage{rotating}
\usepackage{epstopdf}
\usepackage{color}
\usepackage{amsmath,amsthm}
\usepackage{wrapfig}

\begin{document}

\title
{Generalized Iterative Annealing Model for the action of RNA chaperones}

\author{Changbong Hyeon$^{1}$}
\thanks{To whom correspondence should be addressed : hyeoncb@kias.re.kr}
\author{D. Thirumalai$^{1,2}$}
\affiliation{
$^1$School of Computational Sciences, Korea Institute for Advanced Study, Seoul 130-722, Korea\\
$^2$Biophysics Program,Institute for Physical Science and Technology, University of Maryland, College Park, MD 20742, USA}

\begin{abstract}
As a consequence of the rugged landscape of RNA molecules their folding is described by the kinetic partitioning mechanism according to which  only a small fraction ($\phi_F$) reaches the folded state while the remaining fraction of molecules is kinetically trapped in misfolded intermediates. The transition from the misfolded states to the native state can far exceed biologically relevant time. Thus, RNA folding {\it in vivo} is often aided by protein cofactors, called RNA chaperones, that can rescue RNAs from a multitude of misfolded structures.   
We consider two models, based on chemical kinetics and chemical master equation,  for describing assisted folding. 
In the  passive model, applicable for class I substrates, transient interactions of misfolded structures with RNA chaperones alone are sufficient to destabilize the misfolded structures, thus entropically lowering the barrier to folding. For this mechanism to be efficient the intermediate ribonucleoprotein (RNP) complex between collapsed RNA and protein cofactor should have optimal stability. We also introduce an active  model (suitable for stringent substrates with small $\phi_F$), which accounts for the recent experimental findings on the action of CYT-19 on the group I intron ribozyme, showing that RNA chaperones does not discriminate between the misfolded and the native states.   In the active model, the RNA chaperone system utilizes chemical energy of ATP hydrolysis to repeatedly bind and release misfolded and folded RNAs, resulting in substantial increase of  yield of the native state. The theory outlined here shows, in accord with experiments, that in the steady state the native state does not form with unit probability. 
\end{abstract}
\maketitle

Since the ground breaking discovery of self-splicing catalytic activity of group I intron ribozymes \cite{CechCell81,DoudnaNature02} numerous and growing list of cellular functions have been shown to be controlled by RNA molecules \cite{Hutvagner01Science,Bartel09Cell}. These discoveries have made it important to determine how RNA molecules fold \cite{Thirum05Biochem,Chen08ARB,Woodson10ARB}, and sometimes switch conformations in response to environmental signals \cite{Serganov13Cell} to execute a wide range of activities from regulation of transcription and translation to catalysis. At a first glance, 
it may appear  that RNA folding is simple because of the potential restriction that the  four different are  paired as demanded by the  Watson-Crick (WC) rule.  However, there are several factors that make  RNA folding  considerably more difficult than the more thoroughly investigated protein folding problem \cite{Thirum05Biochem}. 
The presence of  negative charge on the phosphate group of each nucleotide, participation of a large fraction of nucleotides in non WC base pairing \cite{Dima05JMB}, the nearly homopolymeric nature of purine and pyrimidine bases, and  paucity of structural data  are some of the reasons that render the prediction of RNA structures and their folding  challenging \cite{Thirum05Biochem}. Despite these difficulties considerable progress has been made in understanding how large ribozymes fold  {\it in vitro} \cite{Thirumalai96ACR,Treiber01COSB}. These studies have shown that the folding landscape of RNA is rugged consisting of many easily accessible Competing Basins of Attraction (CBAs) in addition to the Native Basin of Attraction (NBA), which implies that the stability gap \cite{Guo92JCP,Goldstein92PNAS} separating the CBAs and the NBA is modest relative to proteins. 
As a consequence of the rugged folding landscape, only a small fraction of initially unfolded molecules reaches the NBA rapidly while the remaining fraction are kinetically trapped  in a number of favorable alternative low energy misfolded CBAs, as predicted by the Kinetic Partitioning Mechanism (KPM) \cite{Guo95BP}. The free energy barriers separating the CBAs and the NBA is often  high. Consequently, the transitions times to the NBA from the CBAs could exceed biologically relevant time scale ($\mathcal{T}_B$). The upper bound for $\mathcal{T}_B$ should be no greater than tens of minutes given the typical cell cycle time. 

Because of the modest stability gap even simple RNA molecules could misfold at the secondary as well as tertiary structure levels. In structural terms, secondary structure rearrangements, which are observed in the folding of P5abc \cite{Wu98PNAS} and riboswitches \cite{Montange08ARB} induced by metal ions and metabolites, respectively, are one cause of  the high free energy barriers separating CBAs and NBA in RNA.   
The free energy barrier associated with melting of $n$ base pairs is $(n-1)\times f_{bp}$ where $f_{bp}$, the free energy stabilizing a base pair, is $\sim$ $(2 - 3)$ kcal/mol/bp \cite{MathewsJMB99}. The average length of a duplex in RNA structure is estimated to be $\langle n\rangle\approx 6$ bp from the ratio of nucleotides participating in the duplex formation $f=\frac{2\langle n\rangle}{l_{ss}+2\langle n\rangle}\approx 0.54$ \cite{Dima05JMB, fang2011JPCB}, where $l_{ss}(\approx 6.5)$ \cite{fang2011JPCB} is the average length of a single stranded chain in native RNAs. Using these estimates, we surmise that  a typical free energy barrier associated with secondary structure rearrangement is $\delta G^{\ddagger}\approx$ $(10-15)$ kcal/mol ($\approx$ $(17-25)$ $k_BT$). By assuming that the prefactor for barrier crossing is $\tau_o\sim$1 $\mu sec$ \cite{Hyeon12BJ} the time scale for spontaneous melting of a hairpin stack could be as large as $\tau_oe^{\delta G^{\ddagger}/k_BT}\approx (24-10^5)$ sec $\sim$ 1 day! Indeed, several {\it in vitro} experiments have shown that {\it Tettrahymena} ribozyme does not reach the folded state with unit probability even after hundreds of minutes \cite{Pan97JMB}. The sluggish RNA folding kinetics {\it in vitro} is reminiscent of that observed in glasses due to the presence of multiple metastable states (CBAs) \cite{Kirkpatrick89JPhysA}.
Because of trapping in long-lived CBAs, it is practically impossible for a large ribozyme to spontaneously make a transition to the native state with substantial probability within $\mathcal{T}_B$.   These considerations suggest  \textit{in vivo} folding would require RNA chaperones \cite{Herschlag95JBC,Thirum05Biochem}.

The goal of this paper is to produce a quantitative framework for understanding the function of RNA chaperones, which are protein cofactors that interact with the conformations in the CBAs and facilitate their folding. We classify RNA chaperones as passive and active. Passive chaperones transiently interact with RNA molecules and reduce the entropy barrier to folding without requiring an energy source. On the other hand, active chaperones function most efficiently by lavish consumption of ATP in the presence of DEAD-box proteins.  The need for passive or active chaperones depends on the client molecules and the extent of misfolding (see below). We formulate a general kinetic model to describe both passive (no ATP required) and active (requires ATP hydrolysis) roles RNA chaperones play in rescuing misfolded states. The resulting theory accounts for experimental observations, and should be useful in quantitatively analyzing future experiments.

\section*{Classification of RNA substrates}
The principal role of chaperones is to assist in the resolution of the
multitude of alternative misfolded structures that RNA readily adopts so that sufficient
yield of the native material is realized in biologically viable
time less than $\mathcal{T}_B$. Because spontaneous yield of the native state of large ribozymes even at high Mg$^2+$ concentrations is small \cite{Pan97JMB}, it is likely that {\it in vivo} RNA chaperones are required to boost the probability of reaching the folded state within  $\mathcal{T}_B$. Unlike the well-studied bacterial GroEL-GroES, a well-identified ``one-fit-all'' chaperonin system for processing cytosolic proteins \cite{ThirumalaiARBBS01}, protein-cofactors that act as RNA chaperones vary from one RNA to the other \cite{moll2003EMBOreport,Lorsch2002Cell,SchroederGD02}. Based on a number of experiments (see \cite{Thirum05Biochem,Woodson10RNABiol} for reviews)
we classify the client RNA molecules into two classes depending on the need for the RNA chaperones to utilize the free energy of ATP hydrolysis in facilitating folding. 

\begin{enumerate}
\item	Folding of  class I RNA molecules is greatly aided by the interactions with protein cofactors while their assistance may  not be strictly required. These RNA molecules are not stringent RNA substrates.
For example, the splicing reaction of mitochondrial bI5 group I intron is activated in 50 mM or greater Mg$^{2+}$ concentration at room temperature but interactions with cytochrome b pre-mRNA processing protein 2 (CBP2) or {\it Neurospora Crassa} mitochondrial tyrosyl tRNA synthetase (CYT-18) enables  splicing at physiological level ($\sim$7mM) of Mg$^{2+}$ by enhancing  folding of the bI5 core. 

\item \emph{Tetrahymena} ribozyme and other group I introns belong to stringent class II substrates. Spontaneous folding, even at high counterion concentration, occurs too slowly with low yield of the native state  to be biologically viable. 
At high temperatures  folding of the misfolded \emph{Tetrahymena} ribozyme is aided by  formation of ribonucleoprotein (RNP) assembly with the promiscuously interacting CYT-18, which in essence follows the mechanism of passive assistance \cite{Thirum05Biochem}. However, an ATP-dependent helicase activity associated  CYT-19 produces functionally competent states that can splice efficiently at normal growth temperature.  Although not firmly established, it is suspected that RNA chaperones bind to single stranded regions of the misfolded structures, which upon release places the RNA in a different region of  the folding landscape, giving it a new opportunity to fold just as anticipated by the Iterative Annealing Model (IAM) \cite{Todd96PNAS}.
\end{enumerate}

The fundamental difference between class I and class II RNA substrates is in the apparent time scale of catalysis ($\tau^{app}_{cat}$) by the ribonucleprotein (RNP) complex formed between RNA and the RNA chaperone. 
If this time scale is smaller than $\mathcal{T}_B$ ($\tau_{cat}^{app}\ll \mathcal{T}_B$), the formation of RNP alone is sufficient to produce functionally competent RNA molecules. 
In the opposite case ($\tau_{cat}^{app}\gg \mathcal{T}_B$), the conversion of misfolded RNA into folding competent form needs assistance from a specially designed action of RNA chaperone that can transduce the free energy of ATP hydrolysis. Below we will describe a mathematical model for the two scenarios.

\section*{Passive assistance and the Tertiary Capture Model}
\begin {figure}[ht]
\centering \includegraphics[width=5.0in]{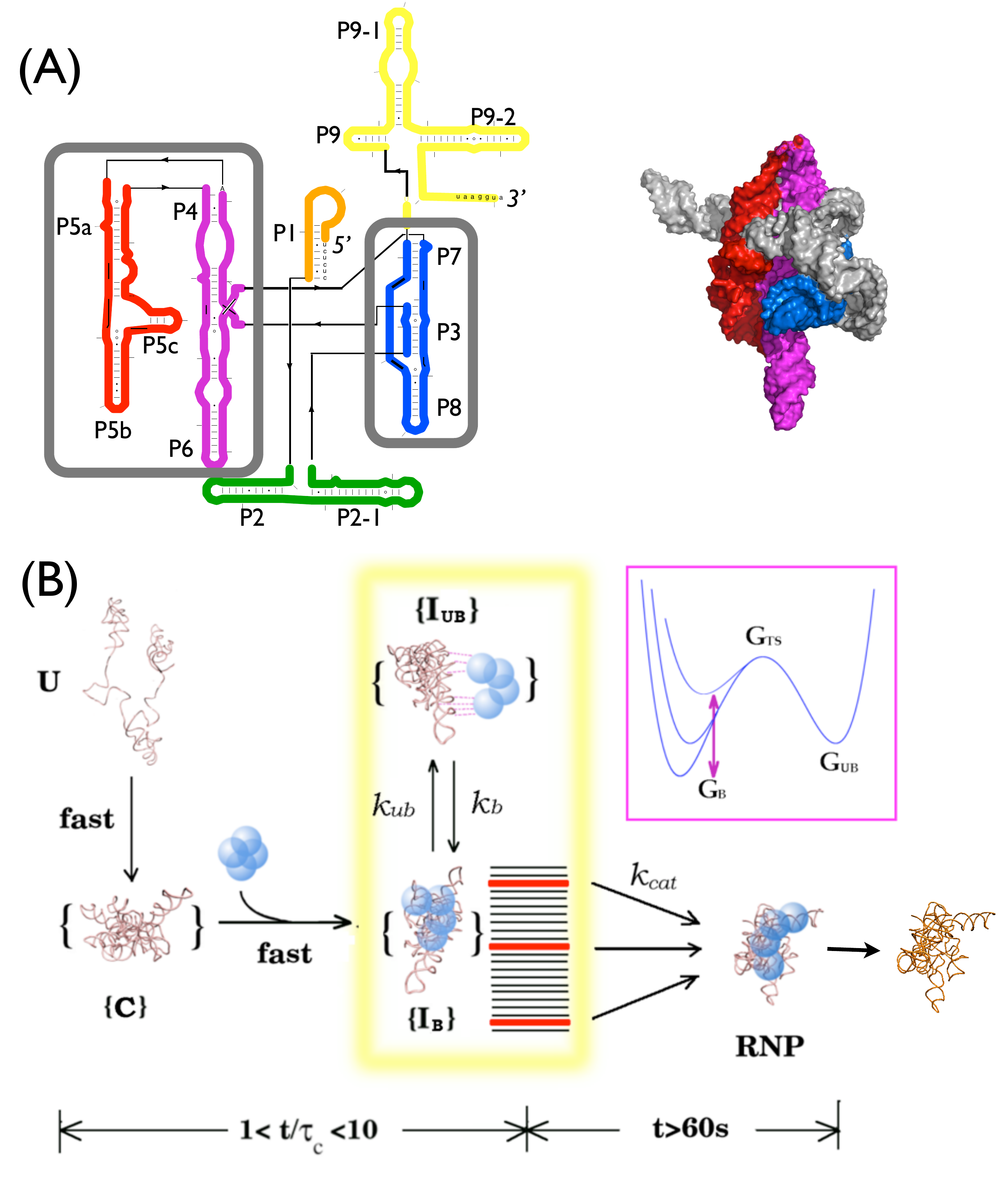}
\caption{(A) On the left is shown the secondary structure map of {\it Tetrahymena} ribozyme, a natural substrate for RNA chaperones. The schematic structure on the right is the three dimensional structure of \emph{Tetrahymena} ribozyme with P4-P5-P6 and P7-P3-P8 domains highlighted. (B) A model for passive assistance of RNA folding by protein cofactor, which generalizes tertiary capture model proposed by Garcia and Weeks \cite{garcia2004structural}. 
Rapidly collapsed RNA intermediate upon addition of multivalent counterions forms with a protein cofactor (blue spheres) an intermediate ensemble of RNP complex, which fluctuates between $\{I_B\}$ and $\{I_{UB}\}$. 
$\{I_{UB}\}$ is an ensemble of transiently unbound intermediate. 
Binding and transient release anneal the misfolded RNA, leading to the assembly competent RNP containing the folded RNA. The final step is the release of the native RNA.
\label{fig:passive_RNP}
} 
\end{figure}

The tertiary structure capture model \cite{weeks1996science,WeeksCOSB97,buchmueller2000NSMB,garcia2004structural} accounts for the passive action of RNA chaperones in the folding of mitochondrial bI5 group I intron without ATP. 
Explicit mechanisms of recognition by passive RNA chaperones by the collapsed RNAs may differ for different systems, and might also depend on whether the RNA collapse is specific or non-specific. In a majority of cases ribozymes undergo an extended to a collapsed transition even at modest ion concentration producing a heterogeneous population of compact structures whose affinity for the protein cofactors could vary greatly. For example,  
 CBP2 could bind to these compact structures with partially folded cores (P5-P4-P6 and P3-P7-P8) of group I intron (Fig.\ref{fig:passive_RNP}A) with differing specificity, and promote the subsequent assembly of 5' domain of bI5 core. In contrast, CYT-18 binds to RNA and forms a stable CYT-18-bI5 complex at an early stage of RNA folding and promotes the splicing competent states \cite{Webb01JMB}. If the association between the cofactor and compact RNA is too weak then large conformational fluctuations can produce long-lived entropically stabilized metastable kinetic traps for RNA. In this case, the protein cofactor would have little effect on RNA folding. In the opposite limit, when the cofactor interacts strongly with collapsed RNA, transient unfolding in the RNA conformations, which are needed for resolving misfolded structures to the native state, would be prohibited. Thus,
for the chaperone-assisted folding of class I RNA substrates, an optimal stability of the RNA-cofactor intermediate is needed to efficiently produce an assembly-competent RNP complex.

The physical picture of passive assistance of RNA chaperones described above, encapsulated in the Weeks-Cech tertiary capture mechanism,  can be translated  into the kinetic scheme shown in Fig.\ref{fig:passive_RNP}. 
After RNA collapses rapidly to an ensemble of collapsed intermediate structures \{C\} ($U\rightarrow \{C\}$) consisting of a mixture of specifically and non-specifically collapsed structures, promiscuous binding of chaperone (blue spheres in \ref{fig:passive_RNP}) to the conformations in  \{C\} produces a fluctuating ensemble of tightly and loosely bound intermediate RNP complex $\{I_B\}\rightleftarrows\{I_{UB}\}$. This process is conceptually similar to the encounter complex in protein-protein interaction \cite{Froloff97ProteinScience,Clore07Nature,lee2013real}.  
Only a fraction ($\phi_F$) of states among the tightly bound ensemble of RNP, $\{I_B\}$, is viable for producing functionally competent RNP state.  Thus, $\{I_B\}$ is partitioned roughly into  
$\{I_B\}=\{I_B^{cat}\}+\{I_B^{nc}\}$, where $\{I_B^{cat}\}$ denotes the intermediate ensemble that can fold into the competent RNP while $\{I_B^{nc}\}$ cannot. 
Since transitions among the states in $\{I_B\}$ is non-permissible on viable time scales, the only way for a molecule trapped in $\{I_B^{nc}\}$ to reach the competent RNP state is to visit a transiently unbound (or loosely bound) intermediate ensemble $\{I_{UB},\}$ and  explore the states belonging to $\{I_B^{cat}\}$.
Once the RNA is in \{$I_B^{cat}$\} ensemble, the rate of RNP formation is given by 
\begin{align}
r=k_{cat}[I_B^{cat}], 
\end{align} 
which can be quantified by assuming  steady state production of $[I_B^{cat}]$, i.e.,  
$d[I_B^{cat}]/dt\approx 0 = -(k_{cat}+k_{ub})[I_B^{cat}]_{ss}+k_b\phi_F[I_{UB}]_{ss}$. 
Defining a constant for rapid pre-equilibration ($k_{cat}\ll k_b$, $k_{ub}$) between the two collapsed intermediate  ensembles 
$K_{eq}^I=[I_{UB}]_{ss}/[I_B]_{ss}=k_{ub}/k_b$
with a total concentration of collapsed intermediate state $I_0=[I_{UB}]+[I_B]$, we obtain the rate of RNP formation at steady state: 
\begin{align}
r_{ss}=\left(\frac{k_{cat}}{k_{cat}/k_b+K_{eq}^I}\right)\left(\frac{K_{eq}^I}{1+K_{eq}^I}\right)\phi_FI_0.
\label{eqn:rss}
\end{align}
A change in the strength of binding between RNA and protein cofactor would affect the values of 
$K_{eq}^I\left(=e^{-(G_{UB}-G_B)/k_BT}\right)$ 
by modulating $G_B$ or the stability of $\{I_B\}$ ensemble, while keeping other rate constants ($k_{cat}$ and $k_b$) unchanged (the inset of Fig.\ref{fig:passive_RNP}). 

It can be argued that there be an optimal stability for $\{I_B\}$ in order to maximize the rate of RNP formation. 
If $\{I_B\}$ is too stable compared with $\{I_{UB}\}$ then $\{I_B\}$ becomes a dead end with negligible probability of reshuffling its population into non-productive ensemble of $\{I_B^{nc}\}$ into $\{I_B^{cat}\}$ through conformational fluctuations. 
In contrast, if $\{I_{UB}\}$ is more stable than $\{I_B\}$ the production of competent RNP would be inefficient.     
It is clear from Eq.\ref{eqn:rss} that the limiting condition of $K_{eq}^I$, either $K_{eq}^I\gg k_{cat}/k_b$ or $K_{eq}^I\ll 1$, leads to a vanishing value of $r_{ss}$; 
hence it follows that there is an optimum value $K^I_{eq}=\tilde{K}_{eq}^I$ that maximizes the rate of RNP production. The maximum rate is obtained using $r_{ss}'(\tilde{K}_{eq}^I)=0$: 
\begin{align}
r_{ss}^{max}=\frac{k_{cat}}{(1+\tilde{K}_{eq}^I)^2}\phi_F I_0
\end{align}
where $\tilde{K}_{eq}^I=\sqrt{k_{cat}/k_b}$. 
The presence of $\tilde{K}_{eq}^I$ that maximizes $r_{ss}$ is indicative of an optimal unbinding rate ($\{I_B\}\rightarrow \{I_{UB}\}$), for the formation of competent RNP, that satisfies $\tilde{k}_{ub}=\sqrt{k_{cat}k_b}$. 
As long as $(r_{ss}^{max})^{-1}$ remains less than the biologically viable time scale $\mathcal{T}_B$ ($(r^{max}_{ss})^{-1}<\mathcal{T}_B$), RNA chaperone promotes RNA molecule to reach the functionally competent form by merely providing a suitable molecular interface on which  RNA could interact and anneal its conformation. Physically, this situation is not that dissimilar to the role mini chaperone (apical domain of GroEL) plays in annealing certain non-stringent substrates \cite{Wang00JMB}.

\section*{Generalized Iterative Annealing Model (IAM) for  RNA chaperones}
If sufficient yield of the folded RNA is not realized  on the time scale $\mathcal{T}_B$, i.e., $(r_{ss}^{max})^{-1}>\mathcal{T}_B$ (the partition factor, $\phi_F$, in the KPM is small), then a more active role including ATP consumption is required to resolve the misfolded states. 
For kinetically trapped misfolded RNA molecules,  transient unfolding of misfolded elements by RNA chaperones is needed to increase the yield of RNA since it provides another chance for refolding into a functional state.   In  experiments involving CYT-18 and the DEAD-box protein CYT-19 on {\it Neurospora crassa} group I intron \cite{Mohr02Cell}, it was shown that the two protein cofactors (CYT-18 and CYT-19) work in a coordinated fashion by utilizing ATP hydrolysis. 
ATP-dependent activity of CYT-19 was  required for efficient splicing at the normal growth temperature (25 $^o$C) while CYT-18 alone could rescue the misfolded RNA at high temperatures. In this sense, the active participation shares features of  many biological processes including motility of molecular motors \cite{Hyeon11BJ}, and steps in signal transduction pathways \cite{blenis1993PNAS,clark1995science,sorkin2002NRMCB}.

In the absence of RNA chaperone, the KPM predicts that the initial pool of unfolded RNA ribozymes are partitioned into folded and misfolded conformations,  described by the following set of rate equations \cite{Tehver08JMB}. 
\begin{align}
U&\mathop{\longrightarrow}^{k_F}_{\phi_F} N\nonumber\\
U&\mathop{\longrightarrow}^{k_F}_{1-\phi_F} M\nonumber\\
M&\mathop{\longrightarrow}^{k_S}N
\label{eqn:basic}
\end{align}
For a given ribozyme concentration $X_0=[U]+[M]+[N]$, a fraction $\phi_F$ of ribozyme folds into the native state directly at a rate $k_F$, and the remaining fraction ($1-\phi_F$) is  misfolded upon undergoing non-specific  collapse transition. 
The solution of Eq.\ref{eqn:basic}, in terms of the probability of not being folded, is given by $P_{U+M}(t)\approx \phi_F e^{-k_Ft}+(1-\phi_F)e^{-k_St}$, which follows from the  KPM \cite{Thirum05Biochem}.  
In the presence of chaperone ($C$), an additional set of equations involving the actions of chaperone on the different species should be included in  Eq.\ref{eqn:basic}: 
\begin{align}
C+M&\mathop{\longrightarrow}^{\lambda_M} CS\nonumber\\
CS&\mathop{\longrightarrow}^{k_{R}} C+U.
\label{eqn:chaperone}
\end{align}
Here, $\lambda_M$ is the rate associated with the capture process of the misfolded RNA by the RNA chaperone producing the RNP, $CS$; $k_R$ is the rate of the release process. 
Note that RNA captured by the chaperone is released in the form of unfolded state, which provides RNA with an another chance to fold \cite{Thirum05Biochem}. 
The corresponding rate equations for Eqs.\ref{eqn:basic} and \ref{eqn:chaperone} are: 
\begin{align}
\frac{d[U]}{dt}&=-k_F[U]+k_R[CS]\nonumber\\
\frac{d[N]}{dt}&=\phi_F k_F[U]+k_S[M]\nonumber\\
\frac{d[CS]}{dt}&=\lambda_M[M][C]-k_R[CS]
\label{eqn:rate}
\end{align}
with $X_0=[U]+[N]+[M]$ and $C_0=[C]+[CS]$. 

In the context of  chaperone-assisted RNA folding, $k_F\gg k_S\approx 0$. 
$X_0\gg C_0$.  Fig.\ref{fig:chaperone_action}A displays a numerical solution for each chemical species ($N$, $M$, $U$, $C$, $CS$) starting from an initial condition of $[N]=0$, $[U]=1$, $[C]=0.1$. 
Note that all the states in misfolded ensemble are converted to the native state through the reactions 
$C+M\xrightarrow{\lambda_M} CS\xrightarrow{k_R}C+U$, followed by 
$U\xrightarrow[\phi_F]{k_F}N$, resulting in  100 \% yield of the native state if the chaperone cycle is iterated multiple times. 
If there are  $N_c$ iterations, the yield of the native state ($Y$) becomes  $Y=1-(1-\phi_F)^{N_c}$; thus the population of the folded state will increase as $Y(t)\sim 1-e^{-\phi_Ft/T^{ATP}_c}$ where $T_c^{ATP}$ is the ATP concentration dependent cycling time of the chaperone. 
For the case of group I intron, with $\phi_F\approx 0.1$ \cite{Pan97JMB},  the  cycle should be iterated $N_c=22$ times to obtain the yield of more than 90 \% ($Y\geq 0.9$).

\section*{Recognition of the native state within the IAM model}
In the IAM for folding of proteins it is assumed that GroEL does not recognize the folded state of proteins \cite{Todd96PNAS} because the recognition sites are sequestered in the folded state. However, Bhaskaran and Russell \cite{Bhaskaran07Nature} have reported that CYT19 can unfold (at least partially) the native state of group I intron. 
The plausible lines of evidence that CYT-19 can interact with the native ribozymes and destabilize it \cite{Bhaskaran07Nature} are: 
(i) Although CYT-19 does not significantly alter the cleavage activity of group I intron ribozymes below 5 mM-Mg$^{2+}$, the ribozymes incubated with CYT19 under  less stabilizing conditions of 1mM Mg$^{2+}$ are accompanied by a reduction in the cleavage efficiency. 
(ii) A finite steady state values of native and misfolded ribozymes are reached in the presence of CYT19  regardless of their initial population. 
(iii) Refolding of  ribozymes kinetically trapped in the CBAs, unfolded by CYT-19 from both native and misfolded species, follows the same pathway that predominates in the absence of CYT-19. 
(iv) Unfolding efficiency of CYT-19 depends on the stability of RNA.     

\begin {figure}[ht]
\centering \includegraphics[width=4.0in]{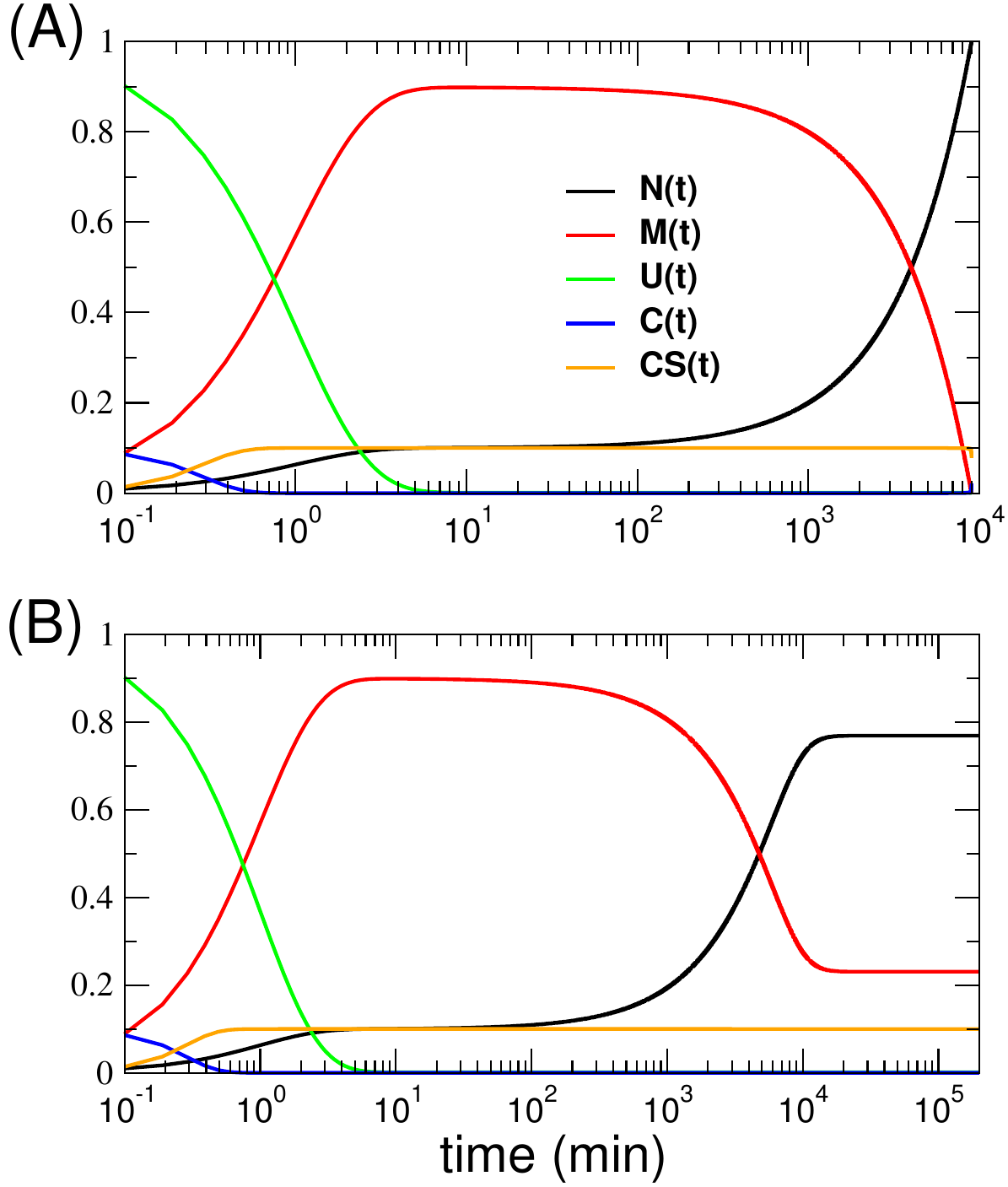}
\caption{Numerical solution of $[N]_t$, $[M]_t$, $[U]_t$, $[C]_t$ and $[CS]_t$ (A) when chaperone does not recognizes the native state, $\lambda_N=0.0$ $min^{-1}$ and (B) when chaperone recognizes the native state, $\lambda_N=1.0$ $min^{-1}$ with $X_0=1.0$, $C_0=0.1$, $[N]_0=0$, $[C]_0=C_0$, $\phi_F=0.1$, $k_F=1$ $min^{-1}$, $k_R=0.01$ $min^{-1}$, $\lambda_M=30.0$ $min^{-1}$. 
\label{fig:chaperone_action}} 
\end{figure}

As indicated by the CYT-19 data \cite{Bhaskaran07Nature}, RNA chaperone can interact with both native and misfolded states under  certain conditions and promote their folding by unfolding (at least partially) them. Incorporation of this finding in our model requires  modification of Eq.\ref{eqn:rate} along the following lines: 
\begin{align}
\frac{d[U]}{dt}&=-k_F[U]+k_R[CS]\nonumber\\
\frac{d[N]}{dt}&=\phi_F k_F[U]-\lambda_N[C][N]\nonumber\\
\frac{d[CS]}{dt}&=(\lambda_N[N]+\lambda_M[M])[C]-k_R[CS]
\label{eqn:general_rate}
\end{align}
 with $X_0=[U]+[N]+[M]$ and $C_0=[C]+[CS]$.
The terms involving $\lambda_N$ account for the recognition of the native state by chaperones ($C+N\xrightarrow{\lambda_N} CS$). 
Typically, $\lambda_N\ll \lambda_M$, implying that the stability  of RNA chaperone in complex with the native state is less than the complex between  misfolded states and the chaperones.
Of particular note is that the numerical solution of Eq.\ref{eqn:general_rate} from the same initial condition used in Fig.\ref{fig:chaperone_action}A shows different behavior. As shown in Fig.\ref{fig:chaperone_action}B, $N$ and $M$ reach steady state values with $[N]\neq 1$ and $[M]\neq 0$.
For a given set of parameters with $\lambda_N\neq0$, the steady state values are:  
\begin{align}
[U]_{ss}&=\frac{k_R}{k_F}[CS]_{ss}\nonumber\\
[N]_{ss}&=\frac{\phi_F k_F}{\lambda_N}\times\frac{[CS]_{ss}}{(C_0-[CS]_{ss})}\nonumber\\
[M]_{ss}&=\frac{(1-\phi_F)k_R}{\lambda_M}\times\frac{[CS]_{ss}}{(C_0-[CS]_{ss})}. 
\label{eqn:steadystate}
\end{align}

\section*{Chemical Master equation formulation for RNA chaperone assisted folding}
The deterministic kinetic schemes presented above do not account for the population of discrete misfolded states explicitly. This can be accomplished by casting the RNA chaperone activity using a chemical master equation (CME) formalism, which implicitly accounts for fluctuation effects due to noise. The corresponding chemical Langevin equation can be derived from the CME under certain approximations \cite{Gillespie00JCP,Zwanzig01JPCB}.  Within the CME inclusion of a finite probability of leakage of flux from the native state facilitated by  RNA chaperones requires a generalization of a formalism developed by previously \cite{Orland97JP} to quantitatively predict GroEL-assisted folding of proteins. In the previous study  it was assumed that GroEL does not recognize the folded state, which does not apply to RNA chaperones. In what follows we include this possibility explicitly within the CME formalism.  

Let us assume that the structural ensemble of RNA can exist in a number of discrete free energy states $f_a$ $(a=0,1,2,\ldots N_c)$.  We assume that $a=0$ is native and all other states with $a\neq 0$ are non-native or misfolded corresponding to the CBAs. 
The occupation probability of each state $P_a(t)$  obeys the following master equation: 
\begin{equation}
\frac{d}{dt}P_a(t)=\sum_bW_{a\leftarrow b}P_b(t)-\sum_bW_{b\leftarrow a}P_a(t).
\label{eqn:master}
\end{equation}

(i) To simplify the above master equation greatly, we formulate the CME for the native state as
\begin{equation}
\frac{d}{dt}P_0(t)=\sum_{b}W_{0\leftarrow b}P_b(t)-\sum_{b}W_{b\leftarrow 0}P_0(t).
\label{eqn:simple_master}
\end{equation}
Assuming that the folding transition rates from the $N_c$ misfolded states to the native state are all similar leads to $\sum_bW_{0\leftarrow b}P_b(t)\approx N_ck_fP_b(t)$; and the transition from the native to misfolded states is $\sum_bW_{b\leftarrow 0}P_0(t)\approx k_uP_0(t)$. 
With $N_cP_b(t)\approx 1-P_0(t)$, the steady state solution of Eq.\ref{eqn:simple_master} is given by   
\begin{equation}
P_0(t)=\frac{k_{f}}{k_{f}+k_{u}}\left[1-e^{-(k_f+k_u)t}\right]. 
\label{eqn:ss_solution}
\end{equation} 
Note that when RNA chaperone recognizes the native state 100 \% yield of the folded state is not achieved even at $t\rightarrow \infty$ and $P_0(\infty)=\frac{k_f}{k_f+k_u}\neq 1$.   
At an irreversible limit $k_f\gg k_u$, which is typically assumed in GroEL assisted protein folding, the yield of native state increases as $P_0(t)\approx 1-e^{-k_ft}$, consistent with the experimental finding \cite{Todd96PNAS}.

(ii) To develop a more sophisticated solution, we assume that detailed balance is satisfied so that $W_{a\leftarrow b}e^{-\beta f_b}=W_{b\leftarrow a}e^{-\beta f_a}$ with $\beta=1/k_BT$ at equilibrium.   
Because the transition rate between two states depends on the height of free energy barrier,  it is not possible to uniquely determine the transition rate from the free energy of each state alone. Nevertheless, for concreteness we  ``choose" $W_{a\leftarrow b}$ as 
\begin{align}
W_{a\leftarrow b}=\tau_0^{-1}\exp{\left[-\frac{\beta}{2}\left(f_a-f_b\right)\right]}. 
\label{eqn:detailed_balance}
\end{align}
In Eq.\ref{eqn:detailed_balance} $\tau_0$ sets the timescale of the problem. 
The master equation for a state $a$ can be written as 
\begin{equation}
\frac{d}{dt}P_a(t)=\sum_{{\text all}\ b}W_{a\leftarrow b}P_b(t)-\sum_{{\text all}\ b}W_{b\leftarrow a}P_a(t).
\label{eqn:other_master}
\end{equation}
By inserting Eq.\ref{eqn:detailed_balance} in Eq.\ref{eqn:other_master}, we get 
\begin{align}
\frac{d}{dt}P_a(t)&=\tau_0^{-1}\left[\sum_{{\text all}\ b}e^{-\frac{\beta}{2}(f_a-f_b)}P_b(t)-\sum_{{\text all}\ b}e^{-\frac{\beta}{2}(f_b-f_a)}P_a(t)\right]\nonumber\\
&=\tau_0^{-1}\left[A_a^{-1}P_{T}(t)-\lambda A_aP_a(t)\right]
\end{align}
where $A_a=e^{\beta f_a/2}$, $P_{T}(t)=\sum_{{\text all}\ b}e^{\beta f_b/2}P_b(t)$, and $\lambda=\sum_{{\text all}\ b}e^{-\beta f_b/2}$. 
The formal solution of this ODE is:  
\begin{align}
P_a(t)=e^{-(\lambda A_a)t/\tau_0}C_a+\tau_0^{-1}A_a^{-1}\int^t_0dt' e^{-\lambda A_a(t-t')/\tau_0}P_{T}(t')
\end{align}
where $C_a=P_a(0)$, the initial probability for state $a$. 
In Laplace domain, $\tilde{P}_a(s)=\int^{\infty}_0dte^{-st}P_a(t)$, 
\begin{align}
\tilde{P}_a(s)=\frac{C_a\tau_0}{z+\lambda A_a}+\frac{A_a^{-1}}{z+\lambda A_a}\tilde{P}_T(s)
\label{eqn:other_Laplace}
\end{align}
where $z\equiv s\tau_0$. 
Insertion of Eq.\ref{eqn:other_Laplace} into $\tilde{P}_T(s)=\sum_{a}A_a\tilde{P}_a(s)$ and rearrangement with respect to $\tilde{P}_T(s)$ leads to 
\begin{align}
\tilde{P}_T(s)=\frac{\sum_{{\text all}\ a}\frac{C_aA_a\tau_0}{z+\lambda A_a}}{1-\sum_{{\text all}\ a}\frac{1}{z+\lambda A_a}}. 
\end{align}
Thus, the probability of native state ($a=0$) in Laplace domain is 
\begin{align}
\tilde{P}_0(s)=\frac{C_0\tau_0}{z+\lambda A_0}+\frac{A_0^{-1}}{z+\lambda A_0}\left(\frac{1}{1-\sum_{{\text all}\ b}\frac{1}{z+\lambda A_b}}\right)\sum_{{\text all}\ b}\frac{C_bA_b\tau_0}{z+\lambda A_b}. 
\end{align} 
The native probability $P_0(t)$ is obtained by the inverse Laplace transform: 
\begin{equation}
P_0(t)=\frac{1}{2\pi i}\int^{C+i\infty}_{C-i\infty}dse^{st}\tilde{P}_0(s)=\sum_{\alpha}R_{\alpha}e^{z_{\alpha}t/\tau_0}
\end{equation}
where $z_{\alpha}$ and $R_{\alpha}$ are the poles and residues of $\tilde{P}_0(s)$. 

The behavior of $P_0(t)$ is determined by the pole structure of $\tilde{P}_0(s)$ and the corresponding residues $R_{\alpha}$.  The residues change depending on the initial conditions specified by $C_b$ ($b=0,1,\ldots$) satisfying $\sum_{{\text all}\ b}C_b=1$. 
Note that $z=-\lambda A_b$ is not a pole but a regular point of $\tilde{P}_0(s)$ for all $b$ because $\lim_{z\rightarrow -\lambda A_b}\left(1-\sum_{{\text all}\ b}\frac{1}{z+\lambda A_b}\right)^{-1}\sum_{{\text all}\ b}\frac{C_bA_b\tau_0}{z+\lambda A_b}=-C_bA_b\tau_0<\infty$. 
$\tilde{P}_0(s)$ has poles at $z=z_{\alpha}$ that satisfies $\sum_{{\text all}\ b}\frac{1}{z_{\alpha}+\lambda A_b}=1$. 
The pole structure from the algebraic solution of $\sum_{{\text all}\ b}\frac{1}{z+\lambda A_b}=1$ can be visualized by plotting $g(z)=\frac{1}{z+\lambda A_0}+\frac{1}{z+\lambda A_1}+\cdots+\frac{1}{z+\lambda A_{N_c}}$. 
$g(0)=1$ since $\lambda =\sum_{{\text all}\ b}A_b^{-1}$. 
In fact, $z_s=0$ is the largest pole of $\tilde{P}_0(s)$ with its residue $R_s=(\lambda^2A_0^2)^{-1}\mathcal{R}(A_0,A_1,\ldots,A_5)$ where $\mathcal{R}(A_0,A_1,\ldots,A_5)=\lim_{z\rightarrow z_s=0}\frac{z}{1-g(z)}$. 

To be specific, we use a model landscape with multiple minima in Fig.3(B), and calculate $g(z)$ and the time dependent behavior of $P_0(t)$ whose behavior depends on the initial condition $C_b$ ($b=0,1,\ldots5$).  
In Fig.3(C) the poles ($z=z_s$, $z_1$, $\ldots$) due to $(1-g(z))^{-1}$ are found at 
$\cdots<z_3<-\lambda A_2<z_2<-\lambda A_1<z_1<-\lambda A_0<z_s=0$. 
Thus, 
\begin{align}
P_0(t)&=R_s+R_1e^{z_1t/\tau_0}+R_2e^{z_2t/\tau_0}+\cdots+R_5e^{z_5t/\tau_0}\nonumber\\
&\sim R_s+R_1e^{-|z_1|t/\tau_0}\mathop{\longrightarrow}^{t\rightarrow \infty} R_s. 
\end{align}
Here note that $z_{\alpha}<0$ for all $\alpha$ and that $P_0(t)$ converges to $R_s\approx 0.864$ independent of any initial condition specified by $C_b$ ($b=0,1,\ldots5$). 
The sign of $R_1\left(=\lim_{z\rightarrow z_1}(z-z_1)\tilde{P}(s)\right)$ changes depending on the value of $C_b$ (see Fig.3(D)), indicating that the population of the native state  adjusts to the equilibrium value regardless of the initial conditions.  
The non-unity steady state value $R_s\neq 1$ of $P_0(t)$ is consistent with Eq.\ref{eqn:ss_solution}, $P_0(\infty)\sim \frac{k_f}{k_f+k_u}$ for $k_f\gtrsim k_u\neq 0$ as well.   
As long as destabilization of the native state is permitted in the chaperone-assisted folding of RNA, complete recovery of the native population cannot occur at $t\rightarrow \infty$, a result that is not only consistent with  experiments \cite{Bhaskaran07Nature} but also follows from mass action kinetic equations.  
Finally, it is worth noting that the value of $R_s$ is the equilibrium population of the native state ($b=0$), i.e., $R_s=\frac{e^{-\beta f_0}}{\sum_{{\text all}\ b} e^{-\beta f_b}}$, which is also confirmed to be identical to the value 0.864 using the parameter set in Fig.3(B).  
Therefore, the iterative action of the RNA chaperone results in the ``annealing" the ensemble of molecules that is kinetically trapped in a multitude of metastable states to reach \emph{equilibrium} by assisting the system to overcome otherwise insurmountable high free energy barriers separating the CBAs and the NBA (Fig.3(A)). 
Only when the stability of native state is far greater than all other metastable intermediates in the CBAs can molecular chaperones bring the yield of native state close to the unity.

\section*{Concluding Remarks}
The tendency of large RNA molecules to misfold readily {\it in vitro} strongly suggests that RNA chaperones must be involved in assisting their folding under cellular conditions. In this paper we discussed two general mechanisms of  chaperone assisted RNA folding. In both scenarios, interactions with protein cofactors  facilitates RNA to escape from the kinetic traps, in the process  annealing misfolded states into functionally competent folded states. Whether the rescue of substrate RNAs occurs by passive (as envisioned in the tertiary capture model) or active mechanism could depend on the nature of RNAs.  For the class I RNA substrates it suffices that the misfolded conformations in the CBAs interact transiently with the protein cofactors. As long as optimal interaction (not too weak or too strong) between RNA and the protein cofactor is achieved the misfolded states can reach the folded state without requiring ATP. On the other hand, for the more stringent class II substrates ATP hydrolysis in the presence of RNA chaperones is coupled to conformational changes in RNA  places the RNA molecules in a different region of the folding landscape from which it can fold with probability $\phi_F$. By repeating this process multiple times  sufficient yield of the native material is generated. We show using mass action kinetics and chemical master equation that as long as the chaperone system does not discriminate between the folded and misfolded states \cite{Bhaskaran07Nature,Grohman13Science} the total yield of the native fold is less than 100\%, which  accords well with experimental findings. 

It is interesting to estimate the work that a typical RNA chaperone system could perform on the RNA. Given that a typical binding free energy between two macromolecules in the cell is $\Delta G\gtrsim -20$ $k_BT$, which we estimate by using an estimated lower bound of dissociation constant $K_d\gtrsim 1$ nM (calculated based on one bound complex in a {\it E. coli} cell), it could be argued that the upper bound of energy or  mechanical work stabilizing the RNP complex is $W\lesssim 20$ $k_BT$.  In the active model, RNA chaperones can facilitate folding to the native state by performing work on the RNA, thus  redistributing the population of native and misfolded RNAs on the folding landscape \cite{Thirum05Biochem,Bhaskaran07Nature}. Using the typical dimensions of the misfolded RNA ($R \approx 4$ nm) the conformational changes in RNA should generate a force of $f = W/R \approx 20$pN, which is large enough to partially unfold compact RNA molecules \cite{Bustamante03Science}.
It is intriguing to note that the free energy associated with ATP hydrolysis is $\delta G_{ATP}\approx 21-25$ $k_BT$ per one ATP molecule, which  lies at the borderline of the maximum binding free energy that is associated with  typical protein-protein or protein-RNA interactions. 

 The individual monomers in muti-subunit molecular chaperones,  which have their own catalytic sites for ATP hydrolysis, typically   form a ring-like structure or act with other cofactors \cite{Jankowsky11TIBS} to further increase the efficiency of free energy transduction by tightly interacting with the target structure and enhancing the generation of mechanical work.  
It is also worth emphasizing that full conversion of ATP hydrolysis free energy to  mechanical work is not always realized. For some chaperones functioning through an active mechanism, conformational cycle of chaperone could remain futile due to inefficient coupling to the structures of the substrate molecules. As a result,  multiple rounds of chaperone cycle are often needed to convert one misfolded molecule to the folded state. 
The variation in the efficiency of free energy transduction from one specific molecular chaperone to the other should lead to a rescaling of the number of iterations $N_c$ to $\varepsilon_u N_c$ where $\varepsilon_u$ is a machine-dependent efficiency of unwinding (unfolding) of the substrate in one  chaperone cycle. 
Thus, the yield of the native state ($Y$) after $N_c$ cycles should be $Y = 1-(1-\phi_F)^{\varepsilon_uN_c}$, leading to $Y(t)\sim 1-e^{-\phi_F\varepsilon_ut/T_c^{ATP}}$, and the time scale for complete  annealing of misfolded biomolecules due to a single molecular chaperone is estimated to be $(\phi_F\varepsilon_u)^{-1}T_c^{ATP}$.  Theses estimates can be experimentally verified provided the stoichiometry of ATP consumption in the active RNA chaperone machinery is measured.\\

\section*{Acknowledgements} 
We thank the Korea Institute for Advanced Study for providing computing resources.
DT acknowledges support from the National Science Foundation through grant CHE 09-14033.


\clearpage

 \begin {figure}[ht]
 \centering \includegraphics[width=5.2in]{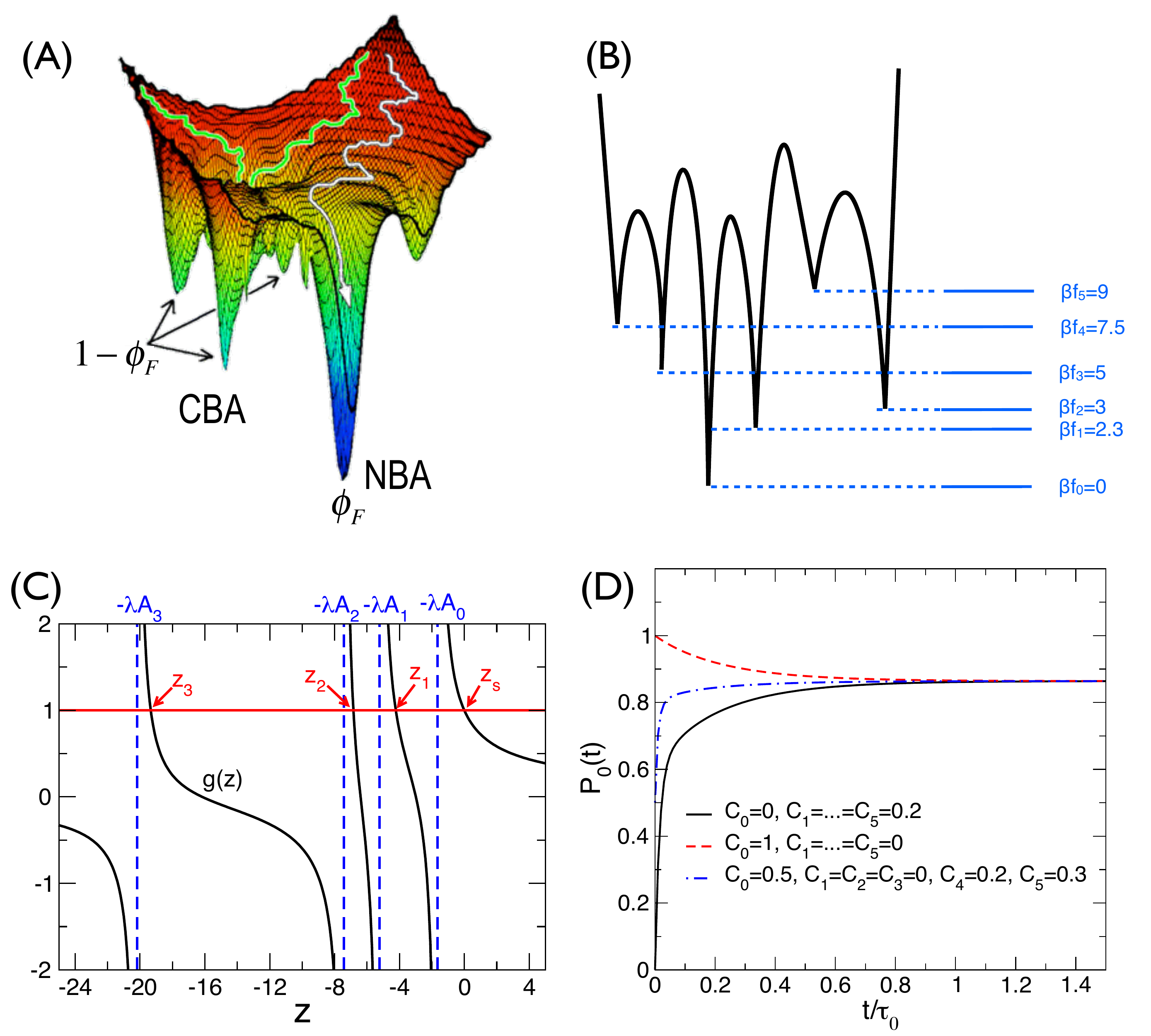}
 \caption{Kinetics of chaperonin assisted RNA folding on a model folding landscape
 (A) Rugged folding landscape of RNA, depicting the native basin of attraction (NBA) and competing basins of attraction (CBA),  and multiple folding routes illustrating the kinetic partitioning mechanism. The figure is adapted from \cite{Thirum05Biochem}. 
 (B) A model folding landscape with 1 native ($b=0$) and 5 misfolded states corresponding to the CBAs in (A) ($b=1,2\ldots,5$). 
The free energy value of each state is assigned as 
 $\beta f_0=0.00$, $\beta f_1=2.30$, $\beta f_2=3.00$, 
$\beta f_3=5.00$, $\beta f_4=7.50$, and $\beta f_5=9.00$, which leads to $\lambda=1.66$. 
(C) The plot of $g(z)=\sum_{{\text all}\ b}\frac{1}{z+\lambda A_b}$ (black solid lines), graphically showing the solutions of $g(z)=1$, i.e., the pole structure of $\tilde{P}_0(s)$ due to $(1-g(z))^{-1}$:   
$z_s=0$, $z_1=-4.22$, $z_2=-6.84$, $z_3=-19.34$, $z_4=-69.49$, $z_5=-148.15$. 
(D) The time evolution of the fraction of native state ($P_0(t)$) with different initial conditions (i) $C_0=0$, $C_1=C_2=\cdots C_5=0.2$ (black, solid line) (ii) $C_0=1$, $C_1=C_2=\cdots=C_5=0$ (red, dashed line) (iii) $C_0=0.5$, $C_1=C_2=\cdots=C_3$, $C_4=0.2$, $C_5=0.3$ (blue, dot-dashed line). 
Note that the fraction of native state in the  steady state is $P_0(t\rightarrow \infty)\neq 1$ due to the flux out of native to non-native states. The steady state value is independent of the initial conditions, suggesting that the RNA chaperones redistribute the population of folded and misfolded states till equilibrium is reached.
\label{fig:OT}
} 
\end{figure}
\end{document}